\documentclass[aps,prl,twocolumn,amsmath,amssymb,superscriptaddress]{revtex4}
\usepackage{epsfig}
\usepackage{graphicx}
\usepackage{dcolumn}
\usepackage{subfigure}
\usepackage{hyperref}
\usepackage{txfonts}
\usepackage{ulem}
\usepackage{bm}
\usepackage{color}
\usepackage{latexsym}
\hypersetup{bookmarks,
            colorlinks,
           citecolor=red,
            filecolor=blue,
           urlcolor=blue,
           pdfpagemode=None}
\usepackage{latexsym}

\def \r{{\mathbf{r}}}

\def \k{{\mathbf{k}}}
\def \q{{\mathbf{q}}}

\def \A{{\mathbf{A}}}
\def \a{{\mathbf{a}}}
\def \K{{\mathbf{K}}}
\def \p{{\mathbf{p}}}
\def \Q{{\mathbf{Q}}}
\def \tn{\textnormal}
\newcommand{\beq}{\begin{equation}}
\newcommand{\eeq}{\end{equation}}
\newcommand{\ba}{\begin{array}{ccc}}
\newcommand{\ea}{\end{array}}

\def\bea{\begin{eqnarray}}
\def\eea{\end{eqnarray}}
\newcommand \eff{\textnormal{eff}}
\begin{document}
\title{Singularity of the London penetration depth at quantum critical points in superconductors}
\author{Debanjan Chowdhury}
\affiliation{Department of Physics, Harvard University, Cambridge, Massachusetts-02138, U.S.A.}
\author{Brian Swingle}
\affiliation{Department of Physics, Harvard University, Cambridge, Massachusetts-02138, U.S.A.}
\author{Erez Berg}
\affiliation{Department of Condensed Matter Physics, Weizmann Institute of Science, Rehovot-76100, Israel}
\author{Subir Sachdev}
\affiliation{Department of Physics, Harvard University, Cambridge, Massachusetts-02138, U.S.A.}
\date{\today}%
\begin{abstract}
We present a general theory of the singularity in the London penetration depth at symmetry-breaking
and topological
quantum critical points within a superconducting phase. 
While the critical exponents, and ratios of amplitudes on the two
sides of the transition are universal, an overall sign depends upon the interplay between the critical theory
and the underlying Fermi surface. We determine these features for critical points to spin density wave and nematic ordering,
and for a topological transition between a superconductor with $\mathbb{Z}_2$ fractionalization and a conventional superconductor.
We note implications for recent measurements of the  London penetration depth in 
BaFe$_2$(As$_{1-x}$P$_x$)$_2$  (Hashimoto {\it et al.\/}, Science {\bf 336}, 1554 (2012)).
\end{abstract}
\maketitle

An important focus of the study of the cuprate high temperature superconductors has been the
quantum criticality of the onset of spin density wave (SDW) order within the superconducting phase.
A number of neutron scattering experiments have observed such critical spin fluctuations in hole-doped
LSCO \cite{keimer1,gabe,boris,chang} and YBCO \cite{keimer2}, and in electron-doped NCCO \cite{greven}.
The phase diagrams of the iron-based superconductors show a clear overlap between the SDW
and superconducting phases \cite{dai}, and Hashimoto {{\it et al.\/} \cite{matsuda12} have recently
provided a careful study of the SDW quantum critical point in BaFe$_2$(As$_{1-x}$P$_x$)$_2$.
A novel feature of the latter observations is
that the influence of the magnetic critical point appears to have 
been observed in a property of the superconducting phase, the London penetration depth.
Such an observation supports the proposal that the `same' electrons are involved in both the SDW order
and superconductivity, and that these two orders are strongly coupled \cite{sachdev12}.

In this paper, we provide a general theory of the singularity in the superfluid stiffness and the London penetration depth 
near a wide class of symmetry-breaking
or topological transitions within superconductors in two spatial dimensions. Our results are summarized in Fig.~\ref{fig:results}
for 3 cases labelled A, B, C.
\begin{figure}
\begin{center}
\includegraphics[width=3.0in]{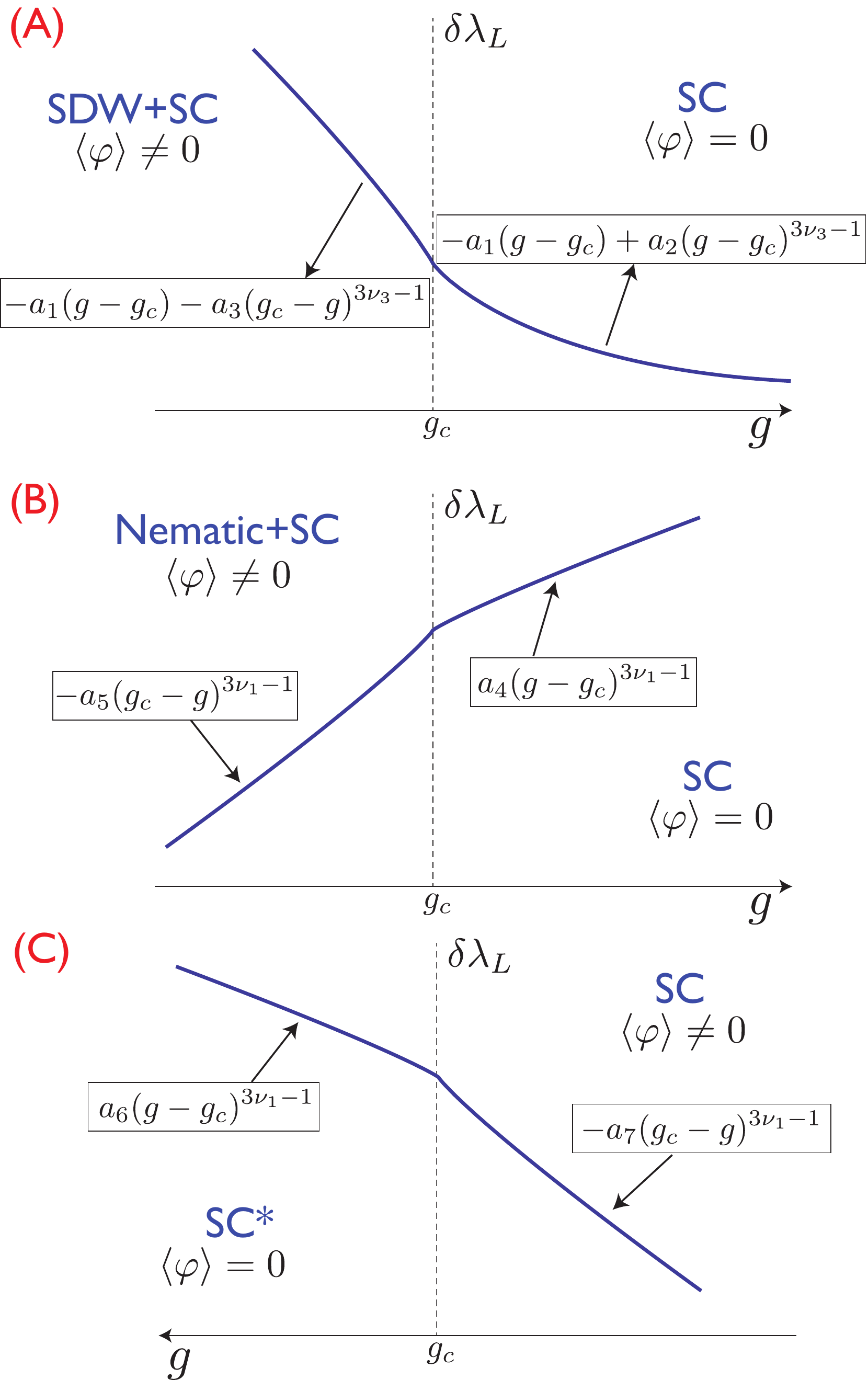}
\end{center}
\caption{Main results for the singular behavior of the London penetration depth, $\lambda_L$ in (A) onset of spin density wave order
leading to transition from SC to SDW+SC, (B) onset of nematic order, and (C) a deconfinement transition to a fractionalized SC* phase.
These transitions are tuned by a generic zero temperature parameter $g$, and the quantum critical point is at $g=g_c$.
The coupling $C_2$ in Eq.~(\ref{Leff}) has values which are (A) negative, (B) positive (or negative, depending on the underlying details of the band-structure), (C) positive.
{\it All\/} the prefactors $a_{1-7}$ are {\it positive}, and some of their ratios are universal:  $a_2/a_3 = 1.52$
and $a_4/a_5 = a_6/a_7 = 0.52$. The correlation length exponents are $\nu_1 = 0.631$ and $\nu_3=0.710$. 
For case A, we have included a non-singular term $- a_1 (g-g_c)$ because it is larger than the singular terms; cases B,C can also
have such non-singular terms, but here they are subdominant to the singular terms.
}
\label{fig:results}
\end{figure}

Case A, the SDW transition, is the one best studied in experiments so far \cite{matsuda12}.
The critical theory is described by the fluctuations of a bosonic SDW order parameter
$\varphi$ with $N=3$ real components, with an effective Lagrangian which has a relativistic form \cite{vicari}; the coupling to the 
fermionic quasiparticle excitations of the superconductor only serves to renormalize the parameters of the Lagrangian. 
These features allow us to use the powerful critical phenomena technology \cite{zinnjustin} to make definitive statements on the singularity in the London penetration depth. We find that the London penetration depth $\lambda_L$ {\it increases} as we approach the quantum critical point from the SC phase.
This is in agreement with the observations \cite{matsuda12}, and an independent recent computation \cite{Lev} which focused on quasiparticle renormalization effects within the SC phase. However, the experiments \cite{matsuda12} also observe a peak-like maximum in $\lambda_L$,
and this is not present in our critical theory for the transition within the superconducting phase. 
This implies that the observed maximum appears {\it within\/} the SDW+SC phase, and not at the
quantum critical point, as has been previously suggested \cite{matsuda12,Lev}.
Explaining the maximum will require consideration of other physical properties of the SDW+SC phase, which
we will briefly discuss at the end of this paper.

Case B applies to the Ising-nematic transition, 
that is also observed in BaFe$_2$(As$_{1-x}$P$_x$)$_2$ \cite{matsudanem}. This transition has a bosonic order parameter $\varphi$
with $N=1$ real component, and the effective theory is otherwise the same as that for the SDW case. However, 
fermionic quasiparticles which are gapless lead to a distinct critical theory \cite{huh}, and so 
our present results apply to the nematic transition only if nodal quasiparticles are absent. 

Finally, case C is a more exotic topological transition between a `fractionalized' superconductor \cite{ssv,piers} (often labeled SC* \cite{TSMPA00})
and a conventional superconductor (SC). Roughly speaking, in a SC* state some of the electrons have localized into a spin liquid
(we consider the case of a $\mathbb{Z}_2$ spin liquid \cite{rs2,wen1}), while the remaining electrons are in a paired BCS state;
there can then be a confinement transition to a SC state which has all the electrons in the BCS state. With the
accumulating evidence for a fractionalized metallic state in a number of heavy fermion 
compounds \cite{canfield,nakatsuji,fried,custers}, we can expect a SC* phase and SC*-SC transition
in the superconducting state at lower temperatures. We will show below that the confinement transition out of the SC* state 
associated with a $\mathbb{Z}_2$ spin liquid is described
by the theory of a ``dual'' Ising field $\varphi$ with a relativistic structure. 
So ultimately, the critical theory is the same as that considered above for the Ising-nematic case, 
with the important difference that it is now the SC phase which has $\left\langle \varphi \right\rangle \neq 0$. 

We now turn to a derivation of the results in Fig.~\ref{fig:results}.
We have already argued that all three cases are described by the familiar $\varphi^4$ field theory 
of a $N$-component field $\varphi$
with imaginary time ($\tau$) action
\beq
\mathcal{S}_\varphi = \frac{1}{2}\int d^2 x d \tau \left[ (\partial_\tau \varphi)^2 + c^2 (\nabla_x \varphi)^2 + 
(g-g_c^0) \varphi^2 + \frac{u}{2} (\varphi^2)^2 \right] \nonumber
\eeq
where $c$ is a velocity of the collective mode excitations of the ordered phase, $g_c^0$ is the bare critical point,
and $u$ is a strongly-relevant self-interaction between $\varphi$ fluctuations. 
The transition is taking place within the superconducting phase, but we can ignore 
the fluctuations of the phase of the superconducting order because these are suppressed by the long-range Coulomb interactions. 

As written, the theory $\mathcal{S}_\varphi$ has no direct coupling to the external magnetic field for cases A and B, and so cannot influence the superfluid 
stiffness and the London penetration depth. In these two cases, the vector potential ${\bf A}$ of the external field couples to the underlying electrons, as does the order parameter $\varphi$.  We will describe below the theory of the electrons coupled to both $\varphi$ and ${\bf A}$, and obtain an effective Lagrangian by integrating out the electronic degrees of freedom. While performing this integration we assume that wavevectors, $q$, of $\varphi$
obey $q \xi_{sc} \ll 1$ where $\xi_{sc}$ is the coherence length of the superconductor. At the same time, $q$ is of order the correlation length, $\xi^{-1}$,
of fluctuations of the bosonic mode $\varphi$; consequently, the validity of our theory is limited to the critical region where $\xi \gg \xi_{sc}$.
With $q \xi_{sc} \ll 1$, we can safely evaluate all gapped fermion loops at $q=0$, and this leads to a simple and local 
effective Lagrangian after the fermions have been accounted for. Moreover, in case C the external magnetic field couples  to the ``dual" Ising field $\varphi$ directly so that in all the three cases, one ends up with,
\beq
\mathcal{L}_{\rm eff} [ {\bf A}, \varphi ] = C_1 {\bf A}^2 + C_2 {\bf A}^2 \varphi^2 , \label{Leff}
\eeq
where $C_{1,2}$ are constants to be evaluated below. 
From this, we obtain the superfluid stiffness as
\beq
\frac{\hbar^2 c^2}{4 e^2} \rho_s = C_1 + C_2 \left\langle \varphi^2 \right \rangle_{\mathcal{S}_\varphi}
\label{rho}
\eeq
where our notation indicates that the expectation value of $\varphi^2$ 
is to be computed in the field theory $\mathcal{S}_\varphi$.
So our final results depend upon two distinct computations.
The first is the computation of $C_2$: we will turn to this below and show how its magnitude and sign depend upon the 
structure of the underlying fermionic excitations. The second is the computation of
$\langle \varphi^2 \rangle $ for which numerous precise
results are readily available \cite{zinnjustin}. Specifically we have
$\langle \varphi^2 \rangle \sim A_\pm |g-g_c|^{3 \nu -1} + \ldots$ as
$g-g_c \rightarrow \pm 0$,
where $\nu$ is the exponent of the correlation length $\xi \sim |g-g_c|^{-\nu}$, the ratio $A_+/A_-$ is universal, 
and the ellipses indicate non-universal terms analytic in $g-g_c$. 
Crucial constraints on the signs of the various coefficients arise from the fact that
$-\partial \langle \varphi^2 \rangle /\partial g$ is proportional to the ``specific heat'', $C_V$,
of the classical statistical system described by the Euclidean field theory $\mathcal{S}_\varphi$,
and so must be positive (note that $C_V$ is unrelated to the specific heat of the quantum model
we are studying). If $\nu > 2/3$, $C_V$ has only a cusp-like singularity at $g=g_c$,
and we assume for this case that $C_V$ is a local maximum at $g=g_c$.
After assembling these constraints with the values of $C_2$ computed below, it is a simple matter
to obtain the results in Fig.~\ref{fig:results}.

We now describe the computation of $C_2$ for case A with SDW order. 
We consider models appropriate for the cuprates and pnictides,
and also the model in Ref.~\cite{sdwsign}, of electrons $c_{i, a\alpha}$ on sites $i$ with orbital index $a$ and spin index $\alpha$
in a SC state described by
\beq
H  = -\sum_{\substack{ij\\ ab}} t_{ab,ij} c_{i,a \alpha}^\dagger c_{j,b\alpha}^{\vphantom\dagger} + \sum_{{\bf k},a} \Delta^a_{{\bf k}} c_{{\bf k},a\uparrow} c_{-{\bf k},a \downarrow} + \mbox{H.c.}
\label{hc}
\eeq
where $t_{ab,ij}$ are the hopping matrix elements, and $\Delta^a_{{\bf k}}$ is the pairing amplitude. 
These electrons are coupled to the 3-component SDW order parameter $\varphi_m$ via
\beq
H_{sdw} = w \sum_{i} \varphi_m (i) \, c_{i,a \alpha}^\dagger \sigma^m_{\alpha\beta} c_{i,b\beta}^{\vphantom\dagger} \, e^{i {\bf K} \cdot {\bf r}_i}
\eeq
where $\sigma^m$ are the Pauli matrices and ${\bf K}$ is the SDW ordering wavevector. We choose the superconducting pairing function
$\Delta^a_{\bf k}$ so that points on the Fermi surface connected by ${\bf K}$ always have pairing amplitudes with opposite sign: this is
the type of spin-singlet superconductivity found in both the cuprates and the pnictides. Finally, we introduce an external magnetic field by a vector potential ${\bf A}$ in the gauge $\nabla \cdot {\bf A} = 0$ via a Peierls substitution in the hopping matrix elements $t_{ij}$. 
Then conventional many-body perturbation theory in the coupling $w$ leads to the Feynman diagrams in 
Fig  \ref{da2} \cite{SI}. It is worth noting that a three-point coupling, $\sim C_3{\bf A}_\perp\varphi^2$, where $\A_\perp$ is the transverse component of $\A$, is also generated upon integrating out the fermions, but $C_3=0$ within the critical region $\xi\gg\xi_{sc}$. We have evaluated $C_2$ numerically for different band-structures that resemble the pnictide and cuprate Fermi surfaces. Remarkably, we find that $C_2<0$ for all the cases that we have considered here; a specific example of a Fermi-surface with the corresponding $\delta\lambda_L$ are shown in Figs.~\ref{fslam} 
and~\ref{lamcup}. In the numerical computations, $\langle\varphi^2\rangle$ has been computed at the Gaussian level on the disordered side of the critical point. Since the qualitative features of the results do not seem to depend on the specific details of the underlying band-structure, it is likely that the observed behavior in $\delta\lambda_L$ is present in the vicinity of all $(2+1)-$dimensional spin-density wave quantum critical points in a superconductor.  We also note that our present methods, which focus only on the longest wavelength fluctuations of $\varphi$, cannot accurately
account for the {\it analytic\/} dependence of $\lambda_L$ on $g-g_c$ contained in the $a_1$ term in Fig.~\ref{fig:results}A;
Ref~\cite{Lev} accounts for the short wavelength fluctuations of $\varphi$ 
more completely, and so their methods give a better estimate of $a_1$, and of the enhancement
of $\lambda_L$ upon approaching the critical point from the SC.

\begin{figure}
\begin{center}
\includegraphics[width=3.4in]{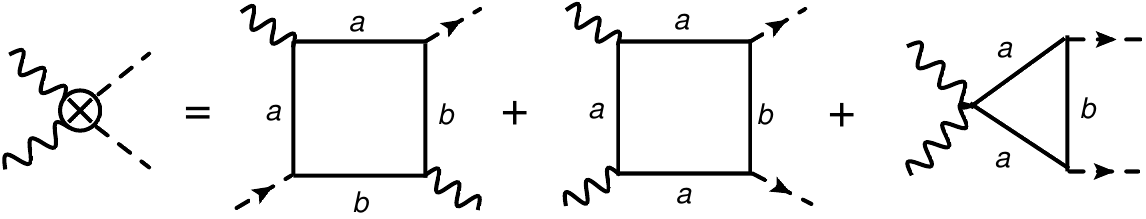}
\end{center}
\caption{The four-point coupling (crossed circle) between ${\bf A}$ and $\varphi$ after integrating out the Fermions. The wavy and the dashed lines represent the gauge field and the $\varphi$ fields respectively. The solid lines represent Fermions having a Nambu structure, with $a\neq b$ for case (A), while $a=b$ for case (B).}
\label{da2}
\end{figure}
A similar analysis applies to the nematic ordering transition in case B. The order parameter $\varphi$ has only one component,
and its coupling to the electrons in the pairing channel is
\beq
H_{{\rm nematic}} = w\varphi\sum_{\k,a} c_{\k,a\uparrow}c_{-\k,a\downarrow}+\tn{H.c.}
\eeq
In this case, the sign of $C_2$ depends on the underlying microscopic details of the band-structure. The behavior of $\delta\lambda_L$ for the Fermi-surface chosen in Fig.\ref{fslam}(A) is shown in the inset of Fig.~\ref{fslam}(B), which corresponds to $C_2>0$. 

\begin{figure}
\begin{center}
\includegraphics[width=3.4in]{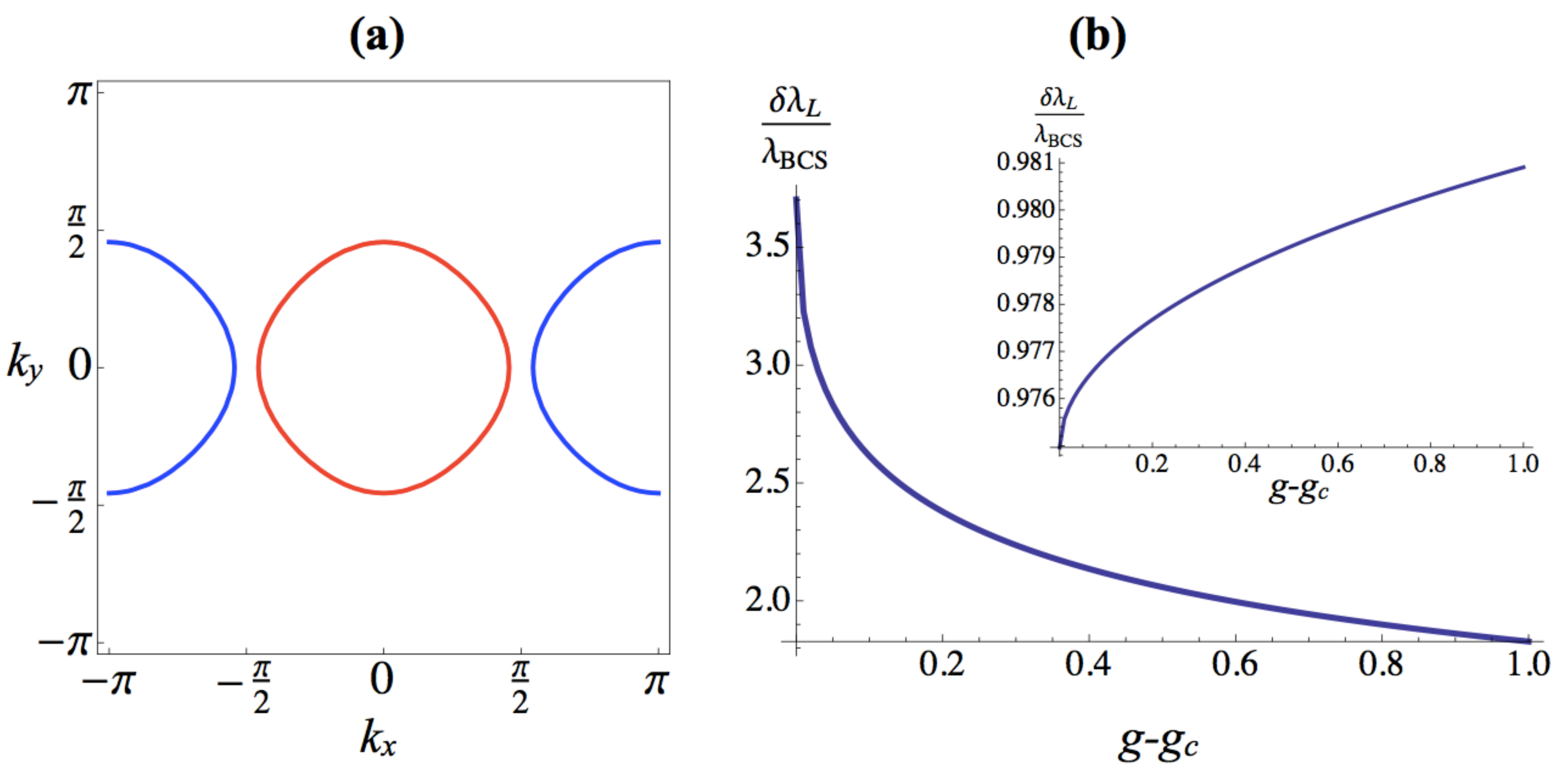}
\end{center}
\caption{(A) A two-orbital band structure of the form $\varepsilon_1(\k)=-2t_1\cos(k_x)-2t_2\cos(k_y)-\mu,~ \varepsilon_2(\k)=2t_2\cos(k_x)-2t_1\cos(k_y)-\mu$, with $t_1=t_2=0.22$, $\mu=-0.5$, $w=1.0$, $\Delta_1=-\Delta_2=1.0$, $c=1$. (B) The singular correction to $\delta\lambda_L/\lambda_{\tn{BCS}}$ as a function of $g>g_c$ for the SDW quantum critical point. Inset: $\delta\lambda_L/\lambda_{\tn{BCS}}$ for $g>g_c$ for the nematic quantum critical point. }
\label{fslam}
\end{figure}
\begin{figure}
\begin{center}
\includegraphics[width=1.0\columnwidth]{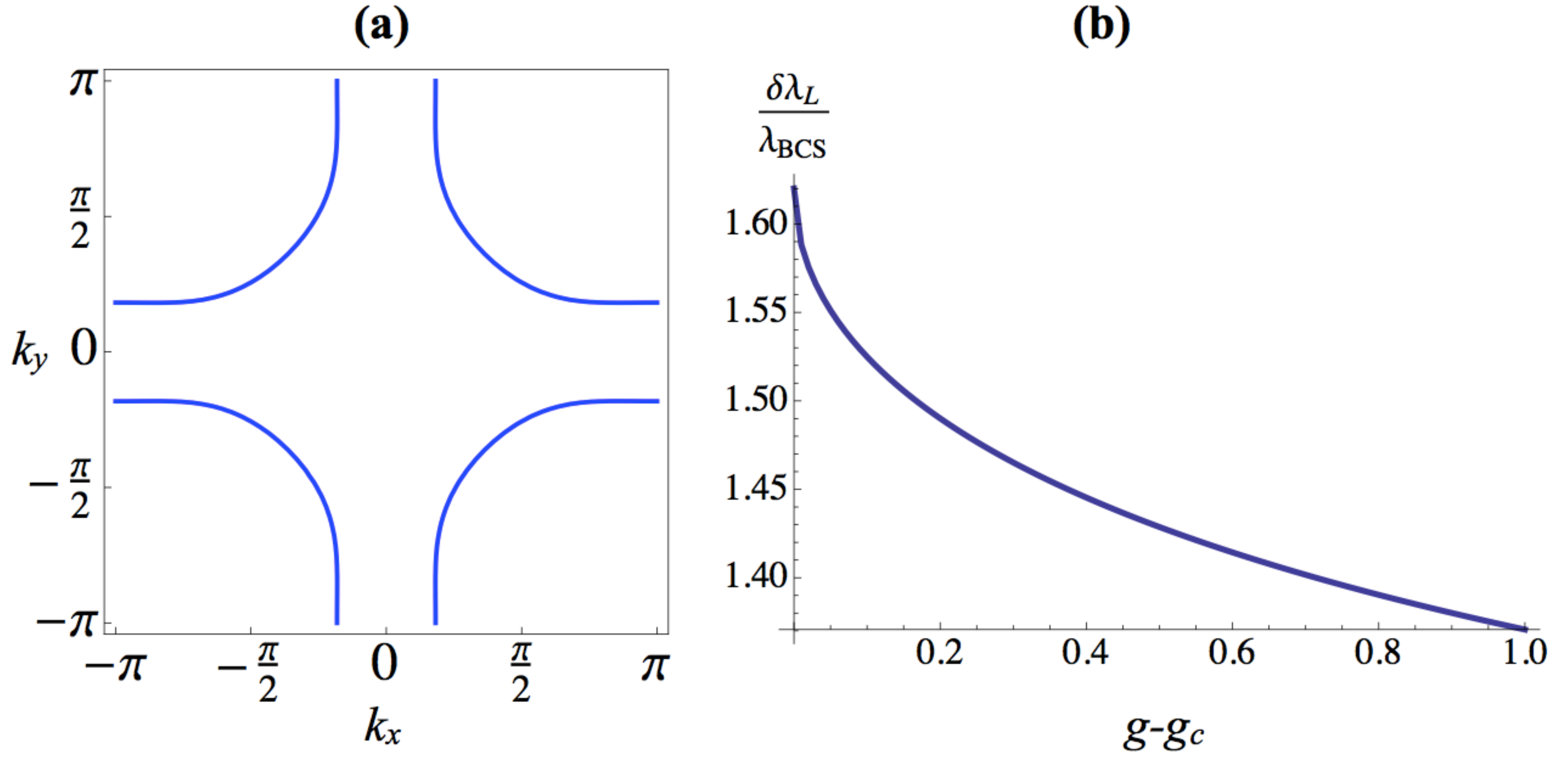}
\end{center}
\caption{(a) Fermi surface for the one-band model, $\varepsilon({\bf k})=-2t_1(\cos(k_x) + \cos(k_y)) - 4t_2\cos(k_x)\cos(k_y) -
 2t_3(\cos(2k_x) + \cos(2k_y))-\mu$, with parameters $t_1=1.0, t_2=-0.32, t_3=0.128, \mu=-1.11856$. The gap function is taken to be $\Delta(\k)=\Delta_0(\cos(k_x)-\cos(k_y))$ with $\Delta_0=1.0$. Other parameters are $ w=1.5, c=1.0$. (b) $\delta\lambda_L/\lambda_{\tn{BCS}}$ is shown for $g>g_c$ for the $\Q=(\pi,\pi)$ SDW QCP.}
\label{lamcup}
\end{figure}

Turning to case C, 
we consider the computation of $C_2$ for the case of the SC*-SC topological transition, where a rather different treatment is required. We consider models appropriate for heavy-fermion materials consisting of itinerant conduction electrons and localized spins. Writing the spin operators in terms of fermionic spinons,   $\vec{S}_i=f^\dagger_{i\alpha}\vec{\sigma}_{\alpha\beta}f_{i\beta}/2$, leads to the emergence of an internal  vector potential, ${\bf a}$, in addition to the electromagnetic gauge field, $\A$. The conduction electrons and the spinons can be described by 
Hamiltonians similar to that in Eq.~(\ref{hc}): for the conduction electrons the pairing amplitude $\Delta$ represents a superconducting pairing,
while the pairing amplitude for the spinons is necessary to obtain a $\mathbb{Z}_2$ spin liquid \cite{rs2,wen1, SI}.
The crucial new ingredient is the Kondo coupling between the conduction electrons and spinons
\beq
H_{\tn {Kondo}}=\sum_{i}(\Phi^*_{i}c_{i\alpha}^\dagger f_{i\alpha}+\Phi_if_{i\alpha}^\dagger c_{i\alpha}),
\eeq
where $\Phi_i$ is a complex field, which carries gauge charge $(-1_\A, 1_\a)$ and whose condensation leads to confinement. Because 
the fermions remain gapped on both sides of the transition, we can integrate them out completely and the action for the gauge fields is given by,
\beq
{\cal{S}}_0[\A,\a] = \int d^2x d\tau \left( \frac{\A_\perp\Pi^c \A_\perp}{2} + \frac{\a_\perp\Pi^f \a_\perp}{2} \right),
\eeq
where $\Pi^{c,f}>0$ denote the  bare ``superfluid'' stiffnesses. 
The condensation of $\langle cc\rangle$ and $\langle ff\rangle$ break the $U(1)_\A$ and $U(1)_\a$ to $Z_{2\A}$ and $Z_{2\a}$ respectively.  The operator $\Phi^2$ is neutral under both $\mathbb{Z}_2$ gauge fields.  However, once $\Phi$ condenses the gauge invariance $\mathbb{Z}_{2A}\times \mathbb{Z}_{2a}$ is broken to its diagonal $\mathbb{Z}_{2(A-a)}$.

We can express $\Phi$ in terms of the real fields $(\varphi_0,\varphi)$. The component $\varphi_0$ remains gapped even across the critical point, and we can safely integrate it out. Therefore, the critical theory is governed by ${\cal{S}}_\varphi$ with $N = 1$. The coupling between $\varphi$ and the gauge-fields is given by \cite{SI},
\beq
{\cal{S}}[\varphi,\A,\a] = \int d^2x d\tau\frac{1}{2}(\A_\perp-\a_\perp)^2\varphi^2.
\eeq
It is now a simple matter to integrate $\a$ out from ${\cal{S}}_0[\A,\a]+{\cal{S}}[\varphi,\A,\a]$, which yields the action in Eq.~(\ref{Leff}) with
\bea
C_1=\frac{\Pi^c}{2},~\tn{and},~ C_2=\frac{\Pi^f}{2(\Pi^f+\varphi^2)}>0.
\eea
In the limit of large $\Pi^f$ and sufficiently close to the critical point, $C_2\approx 1/2$.

This paper has provided  signatures of various quantum critical points in superconductors in Fig.~\ref{fig:results}.
For the important case A of the SDW+SC transition, our result
agrees with observations in BaFe$_2$(As$_{1-x}$P$_x$)$_2$ \cite{matsuda12} in that $\lambda_L$ increases upon 
approaching the critical point from the SC phase. However, one of our key results is that
there is no maximum in $\lambda_L$ at the quantum critical point,
only a change in sign of its second derivative. Therefore the observed maximum should be within the SDW+SC phase,
and involves physics at scales $\xi \sim \xi_c$ or smaller. In this regime, there are
renormalizations of the spectrum and lifetime of the fermionic excitations,
such as those that are crucial for the onset
of the SDW order in a metal \cite{AC,maxsdw}. Therefore, it is necessary to evaluate diagrams like those in Fig.~\ref{da2} while
including Fermi surface reconstruction and fermion self energy corrections.

For case B, there isn't enough evidence yet from the experiments on the pnictides whether the penetration depth has any non-analytic features in the vicinity of the nematic critical point. However, nematic phases are ubiquitous in these systems and hence more careful experiments in the near future are likely to reveal interesting features close to such quantum critical points.  

Finally for case C, the heavy fermion compounds with indications of fractionalized metallic states \cite{canfield,nakatsuji,fried,custers} are good candidates for realizing a SC* superconductor. Moreover,  in CeCu$_2$(Si$_{1-x}$Ge$_x$)$_2$, experiments have found two distinct superconducting domes as a function of pressure alongwith similar results in Ce(Rh,Ir,Co)In$_5$ as a function of doping \cite{yuan}. The nature of one of these SC states remains mysterious and is often attributed to valence fluctuations. It would be interesting to carry out penetration depth measurements in these materials and compare with the present study.

We hope that the signatures reported here will be useful in experimental identifications of various quantum critical points in the pnictides, cuprates and heavy-fermion superconductors in the near future.

{\it Acknowledgements}- We thank A.~Chubukov, J.~Lee, A.~Levchenko, M.~Punk and T.~Senthil for valuable discussions.
This research was supported by the NSF under Grant DMR-1103860, by the U.S. Army Research
Office Award W911NF-12-1-0227, and by a grant from the John Templeton Foundation. B.S. is supported by a Simons Fellowship through Harvard University. E.B. is supported by the ISF under grant no. 7113640101. We thank the Perimeter Institute for hospitality where a part of this paper was written.

\begin{widetext}

\section{Supplementary Material: Singularity of the London penetration depth at quantum critical points
    in superconductors}
In the following sections, we provide additional details about our model and the associated computations. In the next section\ref{sdw}, we discuss our model for the Fermions coupled to a SDW/nematic order at greater length and then provide the necessary details of the diagramatic analysis for the superfluid density. In the last section, we introduce the Kondo-Heisenberg model followed by the low-energy theory and provide a derivation of the important results that were discussed in the main text.
\section{SDW and Nematic QCP: Case (A) and (B)}
\label{sdw}
As introduced in the main text, we start with a band-structure for the Fermions with two orbitals, $c_{i,a\alpha}$. It is convenient to re-write the bands as $c_{\k,1}\rightarrow\psi_{\k,1}$ and $c_{\k,2}\rightarrow\psi_{\k+\K,2}$ \cite{EBMMSS12}. We have therefore shifted the $\psi_2$ fermions by $\K$, the ordering wave-vector; expressing the Yukawa term in terms of $\psi_{1,2}$ is more convenient. The action on a lattice with sites $i,j$ for the SDW problem ($N=3$ component order parameter, $\vec\varphi$) is then given by, ${\cal{S}}=\int_0^\beta d\tau({\cal{L}}_F+{\cal{L}}_{\vec\varphi}+{\cal{L}}_{F\vec\varphi})$, where
\begin{eqnarray}
{\cal{L}}_F&=&\sum_{\substack{i,j\\\alpha, a}}\psi_{i,a\alpha }^\dagger\bigg[(\partial_\tau-\mu)\delta_{ij}-t_{a,ij}e^{ie{\bf A}_{ij}.({\bf r}_i-{\bf r}_j)}\bigg]\psi_{j,a\alpha},\\
{\cal{L}}_\varphi&=&\frac{1}{2}\sum_i\frac{1}{c^2}\bigg(\frac{d\vec\varphi_i}{d\tau}\bigg)^2+\frac{1}{2}\sum_{\langle i,j\rangle}(\vec{\varphi}_i-\vec{\varphi}_j)^2+\sum_{i}\bigg(\frac{g-g_c^0}{2}\vec\varphi_i^2+\frac{u_0}{4}\vec{\varphi}_i^4 \bigg),\\
{\cal{L}}_{F\vec\varphi}&=&w\sum_i\psi_{i,1\alpha}^\dagger(\vec{\sigma}_{\alpha\beta}.\vec{\varphi}_i)\psi_{i,2\beta} + {\textnormal{h.c.}}
\label{lagf}
\label{lagp}
\end{eqnarray}
In the above action, $\mu$ is the chemical potential, $t_{a,ij}$ represent the hopping parameters, $w$ is an ${\it O}(1)$ Yukawa coupling, $\vec{\sigma}$ are the Pauli matrices acting in spin-space and $(g-g_c^0), u_0$ are the bare mass and non-linear self-coupling of the $\vec\varphi$ field, respectively. As before, $a(=1,2)$ represents the orbital label and $\alpha, \beta$ represent the spin labels.  We have introduced an external gauge field, ${\bf A}_{ij}=\sum_{\q}{\bf A}_{\q}~e^{i\q.(\r_i+\r_j)/2}$, defined at the center of each bond $i,j$. We only need terms up to ${\it O}({\bf A}^2)$ in order to compute $\rho_S$. 

Upon expanding ${\cal{L}}_F$ (in the normal state) in powers of ${\bf A}_{ij}$ till second order, we obtain,
\begin{eqnarray}
&&{\cal{L}}_F=\sum_{{\bf k},a,\alpha}[i\omega-\xi_a({\bf k})]\psi_{{\bf k},a\alpha}^\dagger\psi_{{\bf k},a\alpha}
+\sum_{\substack{{\bf k},{\bf q}\\a,\alpha}}{\bf v}_a({\bf k},{\bf q}).{{\bf A}}_{{\bf q}}~\psi^\dagger_{{\bf k}+{\bf q},a\alpha}\psi_{{\bf k},a\alpha }+\sum_{\substack{{\bf q},{\bf q'},{\bf k}\\ a,\alpha, i}}K^i_a({\bf q},{\bf q}',{\bf k})A^i_{{\bf q}}A^i_{{\bf q}'}\psi^\dagger_{{\bf q}+{\bf q}'+{\bf k},a\alpha }\psi_{{\bf k},\alpha  a}+\textnormal{h.c.},
\end{eqnarray}
where $\xi_a({\bf k})=\varepsilon_a({\bf k})-\mu$ and $\varepsilon_a({\bf k})=-\sum_{i,j}t_{a,ij}e^{i{\bf k}.({\bf r}_i-{\bf r}_j)}$ are the dispersions of the orbitals. The paramagnetic current density is given by ${\bf v}_a({\bf k},{\bf q})=e\nabla_k\xi_a({\bf k}+{\bf q}/2)$ and $K^i_a({\bf q},{\bf q}',{\bf k})=e^2\partial_{k_i}^2\xi_a({\bf k}+{\bf q}/2+{\bf q}'/2) / 2$ is the kinetic-energy density along direction $i$. 

As described at the outset in the main text, we shall work in the superconducting state,  with gaps $\Delta_{a}$ on the different orbitals.  Note that the specific material in question has accidental line-nodes on the electron-like sheets, but the nodal excitations don't quantitatively affect our conclusions provided they don't coincide with the hot-spots and couple to the low energy order parameter modes.  We introduce the Nambu spinors, $\Psi^\dagger_{{\bf k},a\alpha}=(\psi^\dagger_{{\bf k},a\alpha}~i\sigma^y_{\alpha\beta}\psi_{-{\bf k},a\beta})$, where $i\sigma^y_{\alpha\beta}=\varepsilon_{\alpha\beta}$. Then,
\begin{eqnarray}
{\cal{H}}_\Delta=\sum_{\k,a\alpha}\Psi^\dagger_{{\bf k},a\alpha}[\xi_a(\k)\tau^z+\Delta_a\tau^x]\Psi_{{\bf k},a\alpha},
\end{eqnarray}
is the modified Hamiltonian for the Fermions in the presence of pairing.  The coupling of the fermions to the SDW ($N=3$, $\vec\varphi$), nematic ($N=1, \varphi$) and gauge fields ($\A$) can be written as,
\begin{eqnarray}
{\cal{L}}_{F\vec\varphi}&=&w\vec{\varphi}.[\Psi^\dagger_{a\alpha}\vec{\sigma}_{\alpha\beta}\hat{\tau}_0\Psi_{b\beta}]\\
{\cal{L}}_{F\varphi}&=&w \varphi\Psi^\dagger_{a\alpha}\hat{\tau}_x\Psi_{a\alpha},\\
{\cal{L}}_{FA}&=&{\bf J}.{\bf A}_{\bf q}+K^i{\bf A}^i_{\bf q}{\bf A}^i_{\bf q'}, ~\textnormal{with}\\
{\bf J}=\sum_{\k, a\alpha}{\bf v}_a(\k,\q)\Psi^\dagger_{\k+\q,a\alpha}\hat\tau_0\Psi_{\k,a\alpha} ~& ,&~
K^i=\sum_{\alpha,\k}K_a^i(\q,\q',\k)\Psi^\dagger_{\k+\q+\q',a\alpha}\hat\tau_z\Psi_{\k,a\alpha}.
\end{eqnarray}
 In the above, $\hat\tau_i$ denote Pauli-matrices acting in Nambu space and $\hat\tau_0=\mathbb{I}_{2\times2}$.
In order to carry out a diagramatic analysis, we work with the Nambu (i.e. normal and anomalous) Green's functions corresponding to the action in ${\cal{L}}_\Delta$. Then the Nambu Green's functions can be written in the following $2\times2$ matrix notation,
\begin{eqnarray}
\hat{G}_a(\p,i\omega_n)&=& \left(\begin{array}{cc}
{\cal{G}}_a^0(\p,i\omega_n)&-{\cal{F}}_a^0(\p,i\omega_n) \\
-{\cal{F}}_a^0(\p,i\omega_n)  & -{\cal{G}}_a^0(-\p,-i\omega_n)\end{array} \right), \textnormal{where}\nonumber\\
{\cal{G}}^0_a({\bf p},i\omega_n)&=&-\frac{i\omega_n+\xi_a({\bf p})}{\omega_n^2+E^2_a({\bf p})} ,~\textnormal{and},~{\cal{F}}^0_a({\bf p},i\omega_n)=-\frac{\Delta_a}{\omega_n^2+E^2_a({\bf p})}.
\end{eqnarray}
In the above, $E_a(\p)=\sqrt{\xi_a^2(\p)+\Delta_a^2}$, $\Delta_1=\Delta,~ \Delta_2=-\Delta$ (for the $s_\pm$ state) and ${\cal{G}}^0, {\cal{F}}^0$ are the normal and anomalous Green's functions.

For the one-band problem with a cuprate-like Fermi surface that we also consider, everything that we have introduced for the two-orbital model goes through with the modification, $f_2(\k)=f_1(\k+\Q)$ (with $\Q=(\pi,\pi)$),where $f_a(...)$ denotes the momentum dependent functions such as the dispersions, velocities, gaps etc. of the orbitals.  Moreover, $\Delta(\k)$ is momentum dependent and has $d-$wave character.

The superfluid density, $\rho_s$, can be inferred from the current-current correlation function by taking the limits of the external frequency and momentum to zero in the correct order \cite{sachdevbook,doug}. When we take ${\bf A}=(A^x,0)$, then the superfluid density is given by,
\begin{eqnarray}
\frac{\rho_s}{\pi e^2}&=&{\cal{K}}_{xx}(q_x=0,q_y\rightarrow0,i\omega_n=0),\\
{\cal{K}}_{xx}({\bf q},0)&=&\frac{\delta^2\ln{\cal{Z}}}{\delta A^x_{\bf q}\delta A^x_{-{\bf q}}}\bigg|_{{\bf A}=0},
\end{eqnarray}
where ${\cal{Z}}=\int{\cal{D}}\vec{\varphi}~e^{-{\cal{S}}_{\textnormal{eff}}}$ is the partition function of the theory corresponding to ${\cal{L}}_{\textnormal{eff}}({\bf A},\vec{\varphi})$ in Eqn.\ref{laga}.
Let us analyze case (A) first, since case (B) proceeds in an almost identical fashion. Upon integrating out the fermions and retaining terms up to second order in {\bf A}, one generates couplings between the vector potential and the SDW field. The resulting action is
\begin{eqnarray}
{\cal{L}}_{\textnormal{eff}}({\bf A},\vec{\varphi})=C_1{\bf A}^2+ C_2\vec{\varphi}^2{\bf A}^2+C_3{\bf A}_\perp\vec{\varphi}^2,
\label{laga}
\end{eqnarray}
where the computation of the expansion coefficients, $C_{1,2,3}$ will be highlighted shortly. 

The bare BCS contribution, $\rho_{\tn{BCS}}$ is given by $C_1$ and can be computed in a standard fashion. The leading singular contribution to the superfluid density at the QCP comes from the other terms above (as shown in Fig. 2 of the main paper). The term corresponding to $C_3{\bf A}_\perp\vec{\varphi}^2$ consists of a ``triangle" graph, but with the paramagnetic current vertex. The dependence on the distance from the critical point, $g-g_c$, will arise once we integrate over $\vec\varphi$, which amounts to closing the dashed lines for the diagrams in $C_2$, while the square of the diagram in $C_3$ would contribute. When the external momentum of the gauge field is zero, these diagrams only depend on the momentum of the SDW fluctuation. However, the fermions are all gapped with a typical energy scale $\Delta$ that is much larger than the SDW energy scale in the vicinity of the QCP.  When the SDW mode goes soft at the critical point, it can't exchange energy/momentum with the massive fermions, which leads the two degrees of freedom to decouple. The couplings then reduce to
 contact interaction, independent of the momentum (and is only determined by the details of the band-structure, gap magnitudes etc.) Therefore, the two sets of diagrams reduce to three- and four-point momentum independent couplings between ${\bf A}$ and $\vec{\varphi}$.

Let us now look at the diagrams in Fig.2---these diagrams effectively give rise to a 4-point coupling between the gauge field and the SDW order parameter. At zero external momentum and frequency, $C_2$ is given by 
\begin{eqnarray}
C_2&=&w^2[I_V+I_\Sigma+I_T], ~~\tn{where}\\
I_V&=&\sum_{{\bf p},ip_n,a\neq b}v_a^x(\p)v_b^x(\p)\tn{Tr}[\hat{G}_a(\p,ip_n)\hat{G}_a(\p,ip_n)\hat{G}_b(\p,ip_n)\hat{G}_b(\p,ip_n)],\\
I_\Sigma&=&2\sum_{{\bf p},ip_n,a\neq b}[v_a^x(\p)]^2\tn{Tr}[\hat{G}_a(\p,ip_n)\hat{G}_a(\p,ip_n)\hat{G}_b(\p,ip_n)\hat{G}_a(\p,ip_n)],\\
I_T&=&2\sum_{{\bf p},ip_n,a\neq b}K_a^x(\p)\tn{Tr}[\hat{G}_a(\p,ip_n)\hat{G}_b(\p,ip_n)\hat{G}_a(\p,ip_n)\hat\tau_z].
\label{c2ex}
\end{eqnarray}
``Tr" stands for trace over the Nambu indices. Upon integrating out the fermions, we also generate a three-point coupling between ${\bf A}$ and $\vec\varphi$. Two of these diagrams combine to give an ${\it O}({\bf A}^2)$ term and is referred to as the {\it Aslamazov-Larkin} diagram. The expression for $C_3$ is given by,
\begin{eqnarray}
C_3=w^2\sum_{{\bf p},ip_n,a\neq b}v_a^x(\p)\tn{Tr}[\hat{G}_a(\p,ip_n)\hat{G}_b(\p,ip_n)\hat{G}_a(\p,ip_n)\hat\tau_z].
\label{c3ex}
\end{eqnarray}
Interestingly, it turns out that $C_3$ is identically zero when evaluated at zero external momentum as it is an odd function of $p_x$ and so we can ignore the three-point coupling between the gauge-fields and $\varphi$. We can compute the traces and carry out the Matsubara sums analytically, leading to expressions that are a function of the lattice momenta. Finally we choose a specific band-structure and evaluate the momentum integrals on a lattice, which gives us a numerical value for $C_2$.

While computing $\lambda$ numerically, we compute $\langle\vec\varphi^2\rangle$ at the Gaussian level, also on the lattice. The bare propagator for the SDW field is given by,
\begin{eqnarray}
D^0({\bf k},i\omega_n)=\frac{1}{\omega_n^2/c^2+[4-2\cos(k_x)-2\cos(k_y)]+g-g_c^0},
\end{eqnarray}
where we have set the lattice spacing to be unity. The bare propagator gets renormalized to $D(\k,i\omega_m)=1/([D^0(\k,i\omega_m]^{-1}-\Pi(\k,i\omega_m))$, where $\Pi(\k,i\omega_m)$ is the polarization bubble due to the superconducting fermions. However the main effect of the gapped fermions is to shift the critical point, $g_c^0$, to a new location, $g_c=g_c^0-\Pi(0,0)$, with some possible renormalization of the spin-wave velocity. Therefore, the functional form of the propagator of the SDW field remains identical to the bare propagator and we shall study the behavior of $\lambda_L$ as a function of the distance away from the renormalized QCP, $g_c$.

The problem is thus reduced to evaluating a one loop effect in the SDW theory with a coupling to external fields obtained from the fermion loops.  Hence the fermions play a crucial but restricted role by determining the sign and size of the coupling.  In particular, the sign of the correction is solely determined by the value of $C_2$ while the shape of the critical bump is determined solely by the SDW theory. The final results are plotted in Figs. 3 \& 4 of the manuscript.

Now the generality of framework can be employed.  When evaluating $\rho_S$ for the nematic problem, we can compute $C_2$, the $4-$point coupling between ${\bf A}$ and $\varphi$, from the same diagrams as in Fig.2, with a few additional differences; one of these includes the fact that the internal $b$ fermion lines are the same as $a$ (there's an overall sum over the different orbitals). The explicit expressions are given by,
\begin{eqnarray}
C_2&=&w^2[I_V+I_\Sigma+I_T], ~~\tn{where}\\
I_V&=&\sum_{{\bf p},ip_n,a}[v_a^x(\p)]^2\tn{Tr}[\hat{G}_a(\p,ip_n)\hat{G}_a(\p,ip_n)\hat{\tau}_x\hat{G}_a(\p,ip_n)\hat{G}_a(\p,ip_n)\hat{\tau}_x],\\
I_\Sigma&=&2\sum_{{\bf p},ip_n,a}[v_a^x(\p)]^2\tn{Tr}[\hat{G}_a(\p,ip_n)\hat{G}_a(\p,ip_n)\hat{\tau}_x\hat{G}_a(\p,ip_n)\hat{\tau}_x\hat{G}_a(\p,ip_n)],\\
I_T&=&2\sum_{{\bf p},ip_n,a}K_a^x(\p)\tn{Tr}[\hat{G}_a(\p,ip_n)\hat{\tau}_x\hat{G}_a(\p,ip_n)\hat{\tau}_x\hat{G}_a(\p,ip_n)\hat\tau_z].
\label{c2ex}
\end{eqnarray}
$C_3$ for this problem is still zero as long as the external momentum/frequency is zero.

\section{SC-$\mathbb{Z}_2$ SC* transition: Case C}
\label{sc*}
Let us start by considering a Kondo-lattice model of conduction electrons ($c$) interacting with localized spins ($f$), described by the following Hamiltonian,
\begin{eqnarray}
H&=&H_c+H_K+H_H, ~\textnormal{where}\\
H_c&=&\sum_{\substack{\langle ij\rangle\\\alpha}}(-t_{ij}e^{ie{\bf A}_{ij}.({\bf r}_i-{\bf r}_j)}-\mu)c_{i\alpha}^\dagger c_{j\alpha}+\sum_{\k}(\Delta^c_\k c^\dagger_{\k\uparrow} c^\dagger_{-\k\downarrow} +\textnormal{h.c.})\nonumber\\
H_K&=&\frac{J_K}{2}\sum_{i}\vec{S}_i.c_{i\alpha}^\dagger\vec{\sigma}^{i}_{\alpha\beta}c_{i\beta}~,~~H_H=J_H\sum_{\langle ij\rangle}\vec{S}_i.\vec{S}_j.
\label{model}
\end{eqnarray}
$H_c$ represents the Hamiltonian of the dispersing conduction electrons that exhibit a superconducting instability with pairing amplitude $\Delta^c$ at low temperatures. $H_K$ represents the Kondo coupling between the conduction electrons and the localized spins. We have included an explicit interaction between the localized spins, $H_H$, which includes effects due to Heisenberg super-exchange and RKKY interactions. ${\bf A}_{ij}$ represents the physical external gauge field.

We shall use the fermionic spinon representation of the localized spins, $\vec{S}_i=f_{i\alpha}^\dagger\vec{\sigma}_{\alpha\beta}f_{i\beta}/2$. Let us now reformulate the above Hamiltonian in terms of a functional integral for the $c$ and $f$ fermions and a slave boson $\Phi\sim\langle c^\dagger f\rangle$. By decoupling $H_K$ and $H_H$ above in terms of Hubbard-Stratanovich fields, we get
\begin{eqnarray}
H'_K&=&\sum_{i}(\Phi^*_{i}c_{i\alpha}^\dagger f_{i\alpha}+\Phi_if_{i\alpha}^\dagger c_{i\alpha})+|\Phi_i|^2/J_K,\\
H'_H&=&\sum_{\langle ij\rangle}(|\chi_{ij}|e^{i{\bf a}_{ij}.(\r_i-\r_j)}f_{i\alpha}^\dagger f_{j\alpha}+\textnormal{h.c.})+|\chi_{ij}|^2/J_H+\sum_{\k}(\Delta^f_\k f_{\k\uparrow}^\dagger f_{-\k\downarrow}^\dagger+\textnormal{h.c.})+|\Delta^f_\k|^2/J_H,
\label{HS}
\end{eqnarray}
where ${\bf a}_{ij}$ is an emergent gauge-field, $|\chi_{ij}|$ is the hopping amplitude for the spinons and we also consider the spinons to be paired with amplitude $\Delta^f$. The complete Hamiltonian of the system is now given by, $H=H_c+H'_K+H'_H+\sum_j\lambda_j(f_{j\alpha}^\dagger f_{j\alpha}-1)$, where $\lambda_j$ is a Lagrange multiplier that restricts the number density of the localized electrons to $n_f=1$ on average. $\Phi_i$ is a complex field, which carries gauge charge $(-1_\A, 1_\a)$ and whose condensation leads to confinement. 
We can now write the following low-energy theory in terms of the gauge-fields $A_\perp, a_\perp$ (transverse components of $\A, \a$) and the scalar field $\Phi$ after having integrated out the fermions (both species of fermions remain gapped on either side of the transition),
\bea
S_{\eff} &=& S_0[\A,\a]+S[\Phi,\tilde{\A}],\\
S_0[\A_\perp,\a_\perp] &=& \int d^2x d\tau \left( \frac{\A_\perp\Pi^c \A_\perp}{2} + \frac{\a_\perp\Pi^f \a_\perp}{2} \right)\\
S[\Phi,\tilde{\A}] &=& \int d^2x d\tau \left(\frac{1}{2} \bigg|\left(\partial - i \tilde{\A}\right)\Phi\bigg|^2 +\frac{U_{\eff}}{2} \right),\\
U_{\eff} &=& b_1 \Phi^2 + b_1^* (\Phi^\dagger)^2 + b_2 \Phi^\dagger \Phi +c_1\Phi^4+c_1^*(\Phi^\dagger)^4+c_2(\Phi^\dagger \Phi)^2+\Phi^\dagger \Phi[c_3\Phi^2+c_3^*(\Phi^\dagger)^2],
\eea
where $\Pi^{c,f}$ denote the bare BCS contribution to the mass for $\A,\a$ due to the $c,f$ Fermions and $\tilde{\A}=\A_\perp-\a_\perp$. By making a particular choice of gauge, we can set $b_1$ to be real ($b_2$ and $c_2$ are also real). Expressing $\Phi$ as $\varphi_0+i\varphi$, with $\varphi_0, \varphi$ as real fields, the most general form of $S[\varphi_0,\varphi,\tilde{\A}]$ is given by,
\begin{eqnarray}
S[\varphi_0,\varphi,\tilde{A}]&=&\int d^2x d\tau \bigg(\frac{1}{2}\bigg[(\partial \varphi)^2 + (\partial \varphi_0)^2\bigg] + U_{\eff}+ \tilde{A}(\varphi\partial \varphi_0 - \varphi_0\partial \varphi) + \frac{1}{2}\tilde{A}^2(\varphi_0^2+\varphi^2) \bigg) ,\nonumber\\
U_{\eff} &=& \frac{r_1}{2} \varphi^2 + \frac{r_2}{2} \varphi_0^2 + \frac{g_1}{4}\varphi_0^4 + \frac{g_2}{4}\varphi^4 + \frac{g_3}{2}\varphi_0^2\varphi^2+g_4\varphi_0^3\varphi+g_5\varphi_0\varphi^3,
\end{eqnarray}
where $r_1=2(b_2-2b_1)$, $r_2=2(b_2+2b_1)$ and all the $g_i$'s are real. We can access the QCP by tuning $r_1$ to zero, where the $\varphi$ mode goes gapless while the $\varphi_0$ mode remains gapped. We can therefore safely integrate $\varphi_0$ out, which leads to a critical theory that is governed by ${\cal{S}}_\varphi$ with $N = 1$, albeit with renormalized coupling constants. The coupling between $\varphi$ and the gauge-fields is given by,
\beq
{\cal{S}}[\varphi,\A,\a] = {\cal{S}}_\varphi + \int d^2x d\tau\frac{1}{2}(\A_\perp-\a_\perp)^2\varphi^2.
\eeq

As we tune across the QCP, there will be a smooth background superfluid density governed by the variation of $\Pi^{c,f}$. However, in this study, we are interested in the singular corrections that arise from the critical fluctuations of $\varphi$ and are dictated by the form of $\langle \varphi^2\rangle$. This is analyzed in the main text in Eqn. (9).

\end{widetext}

\begin{thebibliography}{99}

\bibitem{keimer1} B.~Keimer {\it et al.}, Phys. Rev. B {\bf 46}, 14034 (1992).

\bibitem{gabe} G.~Aeppli, {\it et al.},
Science {\bf 278}, 1432 (1997).

\bibitem{boris} B.~Khaykovich, {\it et al.},
Phys. Rev. B {\bf 71}, 220508 (2005).

\bibitem{chang} J.~Chang, {\it et al.},
Phys. Rev. Lett. {\bf 102}, 177006 (2009).

\bibitem{keimer2} V. Hinkov, {\it et al.},
Materials \& Mechanisms of Superconductivity Conference, Washington D.C. (2012).

\bibitem{greven} E.~M.~Motoyama, {\it et al.},
Nature {\bf 445}, 186 (2007).

\bibitem{dai} P.~Dai, J.~Hu, E.~Dagotto, Nature Phys. {\bf 8}, 709 (2012).

\bibitem{matsuda12} K. Hashimoto {\it et al.\/}, Science {\bf 336}, 1554 (2012).

\bibitem{sachdev12} S. Sachdev, Science, {\bf 336}, 1510 (2012).

\bibitem{vicari} A. Pelissetto, S. Sachdev and E. Vicari, Phys. Rev. Lett. {\bf 101}, 027005 (2008).

\bibitem{zinnjustin} J.~Zinn-Justin, {\it Quantum Field Theory and Critical Phenomena},
(Clarendon Press, Oxford, 2001).

\bibitem{Lev} A. Levchenko, M. Vavilov, M. Khodas and A.V. Chubukov, Phys. Rev. Lett. {\bf 110}, 177003 (2013). Our results for $\delta \lambda_L$ agree with their results on the disordered side of the SDW QCP in case (A). However, they show a maximum in $\lambda$ at the QCP in Fig. (1) of their paper: this was an assumption not supported by any computations on the ordered side. We show here that the actual situation is as in our Fig 1A.

\bibitem{matsudanem} S. Kasahara {\it et al.\/}, Nature {\bf 486}, 382 (2012).

\bibitem{huh}  E.-A. Kim, M. J. Lawler, P. Oreto, S. Sachdev, E. Fradkin, and S. A. Kivelson, Phys. Rev. B {\bf 77}, 
184514 (2008);
Y. Huh and S. Sachdev, Phys. Rev. B {\bf 78}, 064512 (2008).


\bibitem{ssv}  T. Senthil, S. Sachdev and M. Vojta , Phys. Rev. Lett. {\bf 90}, 216403 (2003);
T. Senthil, M. Vojta and S. Sachdev, Phys. Rev. B {\bf 69}, 035111 (2004).


\bibitem{piers} P. Coleman and N. Andrei, J. Phys. Cond. Matt. {\bf 1}, 4057 (1989) and 
N. Andrei and P. Coleman, Phys. Rev. Lett. {\bf 62}, 595 (1989) also considered the interplay between
spin liquids and superconductivity, but do not discuss a phase with the precise characteristics of SC*.


\bibitem{TSMPA00} T. Senthil and M.P.A. Fisher, Phys. Rev. B {\bf 62}, 7850 (2000).

\bibitem{rs2} N. Read and S. Sachdev, 
Phys. Rev. Lett. {\bf 66}, 1773 (1991).

\bibitem{wen1} X.-G. Wen, Phys. Rev. B {\bf 44}, 2664 (1991).

\bibitem{canfield} S.~L.~Bud'ko, E. Morosan, and P.~C.~Canfield, Phys. Rev. B {\bf 69}, 014415 (2004).

\bibitem{nakatsuji} S. Nakatsuji, {\it et al.\/}, Nature Phys. {\bf 4}, 603 (2008).

\bibitem{fried} S. Friedemann, {\it et al.\/}, Nature Phys. {\bf 5}, 465 (2009).

\bibitem{custers} J. Custers, {\it et al.\/},
Phys. Rev. Lett. {\bf 104}, 186402 (2010)


\bibitem{sdwsign} E. Berg, M. A. Metlitski and S. Sachdev, Science {\bf 338}, 1606 (2012).

\bibitem{SI} See Supplemental Material at http://link.aps.org/
supplemental/xxxx for additional details.
\bibitem{AC}
Ar.~Abanov and A.~V.~Chubukov, Phys.\ Rev.\ Lett.\ {\bf 84}, 5608
(2000); {\it Ibid\/} {\bf 93}, 255702 (2004).

\bibitem{maxsdw} M. A. Metlitski and S. Sachdev, Phys. Rev. B {\bf 82}, 075128 (2010); S.~A.~Hartnoll {\it et al.},
Phys. Rev. B {\bf 84}, 125115 (2011).

\bibitem{yuan} H. Q. Yuan et al., Phys. Rev. Lett. {\bf 96}, 047008 (2006); P. G. Pagliuso et al., Physica B {\bf 312-313} 129 (2003).

\end{thebibliography}

\begin{thebibliography}{99}
\bibitem{EBMMSS12} E. Berg, M. Metlitski and S. Sachdev, Science {\bf 338}, 1606 (2012).
\bibitem{sachdevbook} S. Sachdev, {\it Quantum phase transitions}, Cambridge University Press, 2nd Ed. (2011).
\bibitem{doug}  D. J. Scalapino, S. R. White and S. C. Zhang, Phys. Rev. B {\bf 47}, 13 (1993).
\end{thebibliography}
\end{document}